\documentclass{elsart}
\usepackage{ifpdf}
\usepackage{graphicx,natbib,amssymb,lineno}
\ifpdf
\usepackage[%
  pdftitle={Soft and Supersoft X-ray Sources in Symbiotic Stars},%
  pdfauthor={Chunhua Zhu},%
  pdfsubject={The preprint document class elsart},%
  pdfkeywords={X-ray sources, Symbiotic stars},%
  pdfstartview=FitH,%
  bookmarks=true,%
  bookmarksopen=true,%
  breaklinks=true,%
  colorlinks=true,%
  linkcolor=blue,anchorcolor=blue,%
  citecolor=blue,filecolor=blue,%
  menucolor=blue,pagecolor=blue,%
  urlcolor=blue]{hyperref}
\else
\usepackage[%
  breaklinks=true,%
  colorlinks=true,%
  linkcolor=blue,anchorcolor=blue,%
  citecolor=blue,filecolor=blue,%
  menucolor=blue,pagecolor=blue,%
  urlcolor=blue]{hyperref}
\fi

\makeatletter
\def\elsartstyle{%
    \def\normalsize{\@setfontsize\normalsize\@xiipt{14.5}}
    \def\small{\@setfontsize\small\@xipt{13.6}}
    \let\footnotesize=\small
    \def\large{\@setfontsize\large\@xivpt{18}}
    \def\Large{\@setfontsize\Large\@xviipt{22}}
    \skip\@mpfootins = 18\p@ \@plus 2\p@
    \normalsize
}
\@ifundefined{square}{}{\let\Box\square}
\makeatother

\begin{document}

\begin{frontmatter}
\title{Soft and Supersoft X-ray Sources in Symbiotic Stars$^{1}$}

\author{Chunhua Zhu$^{a,b}$,}
\author{Guoliang L\"{u}$^{b}$,}
\author{Zhaojun Wang$^{a,b}$ and}
\author[]{Jun Zhang\corauthref{cor}}{$^{b}$}
\corauth[cor]{Corresponding author.}
\address{$^a$School of Science, Xi'an Jiaotong University, Xi'an, 710049,
China}
\address{$^b$School of Physics, Xinjiang University, Urumqi, 830046,
China} \ead{zhuchunhua@xju.edu.cn(Chunhua Zhu), zhj@xju.edu.cn}
\footnotetext[1]{Supported by the National Science Foundation of
China (Grant Nos. 10647003 and 10763001).}

\begin{abstract}
Assuming that soft X-ray sources in symbiotic stars result from
strong thermonuclear runaways, and supersoft X-ray sources from weak
thermonuclear runaways or steady hydrogen burning symbiotic stars,
we investigate the Galactic soft and supersoft X-ray sources in
symbiotic stars by means of population synthesis. The Galactic
occurrence rates of soft X-ray sources and supersoft X-ray sources
are from $\sim$ 2 to 20 $\rm yr^{-1}$, and $\sim$ 2 to 17 $\rm
yr^{-1}$, respectively. The numbers of X-ray sources in symbiotic
stars range from 2390 to 6120. We simulate the distribution of X-ray
sources over orbital periods, masses and mass-accretion rates of
white dwarfs. The agreement with observations is reasonable.
\end{abstract}

\end{frontmatter}

\section{Introduction}
\label{intro} Symbiotic stars (SySs) are usually interacting
binaries, composed of a cool star, a hot component and a nebula. The
cool component is a red giant which is a first giant branch (FGB) or
an asymptotic giant branch (AGB) star. The hot component is a white
dwarf (WD), a subdwarf, an accreting low-mass main-sequence star, or
a neutron star \citep{kw84,mur91}. The peculiar optical spectrum of
SySs is a very important and interesting phenomenon, and offers some
exciting observational facts \citep{k86,ms99,b00,m03}. \citet{mur97}
showed that the majority of the known galactic SySs are detectable
X-ray sources. According to ROSAT observations of SySs, \cite{mur97}
divided X-ray sources into three distinct classes: $\alpha$-type
X-ray sources in which the photon energies are below 0.4 keV;
$\beta$-type X-ray sources in which the photon energies are roughly
between 0.1 and 10 keV and the peak at about 0.8 keV; and
$\gamma$-type relatively hard X-ray sources like GX 1+4. Following
\cite{v92}, \cite{yun96} and \cite{mur97}, we call $\alpha$-type
X-ray sources as super-soft X-ray sources (SSSs), $\beta$-type X-ray
sources as soft X-sources (SSs)\citep{mur97}. Due to hard X-ray
sources resulting from accreting neutron stars
\citep{v92,mur97,o07}, we do not discuss them in the present paper.

In general, SSSs result from the steady hydrogen burning on
accreting WD in binary systems. While for SSs, it was widely
believed that they are arisen from the colliding winds from the
giant and WD \citep{l95,mur97}. In recent years, many theoretical
studies on SySs have been published, e.g.
\citet{yun95,h95a,it96,h02,lu06,lu07,lu08}. However, there are few
detailed theoretical studies about the Galactic population of X-ray
sources in SySs. \cite{yun96} investigated the Galactic binary SSSs
with white dwarf accretors by means of a population synthesis. They
considered that SSSs in SySs result from the steady hydrogen burning
and thermonuclear runaways, but they did not distinguish SSs from
SSSs.

In the present paper we model possible formation paths for the
Galactic population of SSs and SSSs in SySs with white dwarf
accretors.  In Section 2, we present our assumptions and describe
some details of the algorithm. In Section 3, we discuss the main
results. In Section 4, the main conclusions are given.

\section{The model}
\label{mod} For binary evolution, we use a rapid binary star
evolution (BSE) code of \cite{h02}. Below we describe our algorithm
from several aspects.
\subsection{Symbiotic stars}
SySs are complex binary systems, and can be divided into two
subgroups: `ordinary' SySs, which are assumed to burn hydrogen
steadily; and symbiotic novae, which experience thermonuclear
runaways in their surface hydrogen layers \citep{ty76,m03}. There
are many uncertain physical parameters which can affect the
population of SySs. \citet{lu06} showed that the numbers of SySs and
the occurrences of symbiotic novae are greatly affected by the
algorithm of common envelope evolution, the terminal velocity of
stellar wind $v(\infty)$ and the critical ignition mass $\Delta
M_{\rm crit}^{\rm WD}$ which is necessary mass accreted by WD for a
thermonuclear runaway. The structure factor of the stellar wind
velocity $\alpha_{\rm W}$ and an optically thick wind give a small
uncertainty. In this work, we use the models of SySs in \citet{lu06}
and discuss the Galactic population of SSs and SSSs in SySs.\\

Common envelope:\ Following \cite{nt05} and \citet{lu06}, we use two
algorithms ($\alpha$-algorithm and $\gamma$-algorithm) for common
envelope evolution. The $\alpha$-algorithm results in much shorter
binary separation after undergoing common envelope phase than the
$\gamma$-algorithm. In this work, we take the `combined' parameter
$\alpha_{\rm ce}\lambda_{\rm ce}$ as 0.5 for $\alpha$-algorithm.
$\lambda_{\rm ce}$ is a structure parameter that depends on the
evolutionary stage of the donor. For $\gamma$-algorithm,
$\gamma=1.75$.


$v({\infty})$:\ It is difficult to determine the terminal velocity
of stellar wind $v({\infty})$. \cite{lu06} used two completely
different $v({\infty})$s: \\
(i)$v(\infty)=\frac{1}{2}v_{\rm esc}$, where $v_{\rm esc}$ is the
escape velocity. With ascent of star along giant branch, the stellar
radius becomes larger and larger, which results in the lower
$v(\infty)$.\\
(ii)Using the relation between the mass-loss rates and the terminal
wind velocities fitted by \cite{w03}:
\begin{equation}
\log_{10} (\dot{M}/M_\odot{\rm yr}^{-1})=-7.40+\frac{4}{3}\log_{10}
(v({\infty})/{\rm km \, s^{-1}}). \label{eq:winters}
\end{equation} With ascent of star along giant branch, the
mass-loss rate becomes higher and higher, which results in the
higher $v(\infty)$. However, Eq. (\ref{eq:winters}) is valid for
$\dot{M}$ close to $ 10^{-6}M_\odot$ yr$^{-1}$. For a mass-loss rate
higher than $ 10^{-6}M_\odot$ yr$^{-1}$, Eq. (\ref{eq:winters})
gives too high $v(\infty)$. Based on the models of \citet{win00}, we
assume $v(\infty)=30 {\rm km\, s^{-1}}$ if $v(\infty)\geq30 {\rm
km\, s^{-1}}$.


$\Delta M_{\rm crit}^{\rm WD}$:\ The critical ignition mass of the
nova depends mainly on the mass of accreting WD, its temperature and
material accreting rate. Following \cite{lu06}, we use $\Delta
M_{\rm crit}^{\rm WD}$s by Eq. (A1) of \citet{n04} and Eq. (16) of
\cite{yun95}. As far as whole populations of SySs are considered,
the former gives lower $\Delta M_{\rm crit}^{\rm WD}$ than the
latter \citep{lu06}.


\begin{table*}
  \caption{Parameters of the models for the population of SSs and SSSs in SySs. The first column
           gives the model number. Columns 2, 3 and 4 show the algorithm of the common envelope,
           the terminal velocity and the critical ignition mass, respectively.
           }
  \tabcolsep4.mm
  \begin{center}
  \begin{tabular}{cccc}
  \hline
Cases & Common envelope&$v(\infty)$&$\Delta M_{\rm crit}^{\rm WD}$ \\
Case 1& $\alpha_{\rm ce}\lambda_{\rm ce}=0.5$&$\frac{1}{2}v_{\rm esc}$   &Eq.(A1) in \citet{n04}\\
Case 2& $\gamma=1.75$                        &$\frac{1}{2}v_{\rm esc}$   &Eq.(A1) in \citet{n04}\\
Case 3& $\alpha_{\rm ce}\lambda_{\rm ce}=0.5$&Eq.(\ref{eq:winters})      &Eq.(A1) in \citet{n04}\\
Case 4& $\alpha_{\rm ce}\lambda_{\rm ce}=0.5$&$\frac{1}{2}v_{\rm esc}$   &Eq.(16) in \citet{yun95}\\
\hline
 \label{tab:case}
\end{tabular}
\end{center}
\end{table*}

\subsection{X-ray sources}
 \cite{v92} explained super-soft X-ray
emissions by steady nuclear burning of hydrogen accreted onto WD.
Most of the known SySs are sufficiently bright X-ray sources. As
mentioned in Introduction, \cite{mur97} divided X-ray sources into
SSSs, SSs and relatively hard X-ray resources. It is well known that
the SSSs in SySs are hot WDs whose photospheres can produce
sufficiently hard ($h\nu\approx0.2$kev) photons. For SSs in SySs,
\cite{jor94}, \cite{for95} and \cite{mur97} suggested that they
originate from the collision of two stellar winds. During some
strong thermonuclear runaways, WD can eject some materials with high
velocity ($\sim$ 1000 km s$^{-1}$). Cool giant usually has a high
mass-loss rate with a low velocity (5---30 km s$^{-1}$). Therefore,
a violent collision of the stellar winds is unavoidable in the
strong thermonuclear runaways of SySs. An eruption of the recurrent
nova RS Oph on February 12, 2006 provided the opportunity to perform
comprehensive X-ray observations. \cite{nel08} showed its X-ray
spectroscopy from 0.33 to 10 kev. The spectra indicated a
collisionally dominated plasma with a broad range of temperature and
an energy-dependent velocity structure.  \cite{sok06} and
\cite{orl08} suggested that most of the early X-ray emission of the
2006 outburst of RS Oph originates from the interaction between the
high-velocity ejecta from WD and the circumstellar medium which
results mainly from the stellar wind of the cool giant in SyS.

The high-velocity ejecta is a key factor for producing soft X-ray
emissions in SySs. According to \citet{yar05}, \cite{lu06} divided
the thermonuclear runaways in SySs into two varieties: weak
symbiotic nova in which most of the accreted mater is deposited at
the surface of the WD accretors; strong symbiotic nova in which WD
accretors eject the majority of the accreted mater via high velocity
wind.  We assume SSs in SySs stem from the strong symbiotic novae,
and the steady hydrogen burning `ordinary' SySs and weak symbiotic
novae result in SSSs.

The strength of a thermonuclear runaway depends on the mass and the
mass accretion rate of the WD. Using the results in \cite{yar05},
\cite{lu06} roughly defined the boundary between strong and weak
symbiotic nova via the mass accretion rate and the mass of WD. We
adopt the descriptions in \citet{lu06}. Following \citet{yun96}, we
take the time which takes the WD to decline by 3 mag in its
bolometric luminosity, $t_{\rm 3bol}$, as the lifetime of SSSs. For
SSs which originate from the interaction between the high-velocity
ejecta from WD and the circumstellar medium, we take the duration of
mass loss during strong symbiotic nova, $t_{\rm ml}$, as the their
lifetime. One should note that $t_{\rm 3bol}$ and $t_{\rm ml}$ are
only a zero-order approximation. By a bilinear interpolation
\citep{pre92} of Table 3 in \cite{yar05}, $t_{\rm 3bol}$ and $t_{\rm
ml}$ are calculated from the models in which the temperature of WD
equals $10\times 10^6$ K. If the mass or the mass accretion rate of
the WD in SSs are not in the range of the bilinear interpolation,
they are taken as the most vicinal those.


\subsection{Basic parameters of the Monte Carlo simulation}
We carry out binary population synthesis via Monte Carlo simulation
technique in order to obtain the properties of SSs and SSSs'
population in SySs. For the population synthesis of binary stars,
the main input model parameters are: (i) the initial mass function
(IMF) of the primaries; (ii) the mass-ratio distribution of the
binaries; (iii)
 the distribution of orbital separations; (iv) the eccentricity
distribution; (v) the metallicity $Z$ of the binary systems.

We take IMF in \cite{ktg93} as the primary mass distribution. The
lower and the upper mass cut-offs of the primaries are $0.8M_\odot$
and $8M_\odot$, respectively. For the mass-ratio distribution of
binary systems, we consider only
a constant distribution\citep{kr79,m92,gm94}.

The distribution of separations is given by
\begin{equation}
\log a =5X+1,
\end{equation}
where $X$ is a random variable uniformly distributed in the range
[0,1] and $a$ is in $R_\odot$.

In our work, the metallicity $Z$=0.02 is adopted. We assume that all
binaries have initially circular orbits, and we follow the evolution
of both components by BSE code, including the effect of tides on
binary evolution \citep{h02}.

We assume that one binary with primary mass more massive than 0.8
$M_\odot$ is formed annually in the Galaxy to calculate the
birthrate of SSs and SSSs in SySs \citep{yun94,h95a,h95b}.

\section{Results}
\label{res} We construct a set of models in which we vary different
input parameters relevant to the population of SSSs and SSs in SySs.
Table \ref{tab:case} gives all cases considered in this work. Many
observational evidences showed that the terminal velocity of stellar
wind $v(\infty)$ increases when a star ascends along the AGB
\citep{o02,w03,b05}. In this work, we take $v(\infty)$ calculated by
Eq. (\ref{eq:winters}) as the standard terminal velocity of stellar
wind.

 We take $1\times10^6$ initial binary systems for each case.
For every simulation with $1\times10^6$ binaries, the relative
errors for the symbiotic systems are lower than 1\%. Thus,
$1\times10^6$ initial binaries appear to be an acceptable sample for
our study. The main results of our study are given in Table
\ref{tab:result} and in Figures \ref{fig:numporb} -
\ref{fig:numdmdt}.

\subsection{Population of X-ray sources}
\begin{table*}
  \caption{Different models of the population of SSSs and SSs in
  SySs. The first column gives case number according to Table
  \ref{tab:case}. Columns 2, 3 and 4 give the occurrence rate of SSs which have undergone
  channel I, II and III, respectively. Column 5 shows the total occurrence rate of SSs.
  Columns 6, 7 and 8 give the numbers of SSSs which have undergone
  channel I, II and III, respectively. Column 9 shows the total number of SSSs. In these columns,
  the numbers in parentheses mean SSSs which originate from weak symbiotic novae.
  Columns 10, 11 and 12 show the occurrence rate of SSSs which originate
  from weak symbiotic novae and have undergone
  channel I, II and III, respectively. Column 13 shows the total occurrence rate of
  SSSs which originate from weak symbiotic novae.
           }
  \tabcolsep0.80mm
  \begin{center}
  \label{tab:result}
  \begin{tabular*}{200mm}{|c|c|c|c|c|c|c|c|c|c|c|c|c|}
\cline{1-13} \multicolumn{1}{|c|}{Cases}&\multicolumn{4}{|c|}{SSs
}&\multicolumn{8}{|c|}{SSSs
(weak symbiotic novae or stable hydrogen burning)}\\
\cline{1-13}\multicolumn{1}{|c|}{}&\multicolumn{4}{|c|}{Occurrence
rate($\rm yr^{-1}$)}
&\multicolumn{4}{|c|}{Number}&\multicolumn{4}{|c|}{Occurrence rate  ($\rm yr^{-1}$)}\\
\multicolumn{1}{|c|}{}&\multicolumn{4}{|c|}{(strong symbiotic
novae)}
&\multicolumn{4}{|c|}{}&\multicolumn{4}{|c|}{(weak symbiotic novae)}\\
\cline{1-13}\multicolumn{1}{|c|}{}
&\multicolumn{1}{|c|}{I}&\multicolumn{1}{|c|}{II}&\multicolumn{1}{|c|}{III}&\multicolumn{1}{|c|}{Total}&
\multicolumn{1}{|c|}{I}&\multicolumn{1}{|c|}{II}&\multicolumn{1}{|c|}{III}&\multicolumn{1}{|c|}{Total}
&\multicolumn{1}{|c|}{I}&\multicolumn{1}{|c|}{II}&\multicolumn{1}{|c|}{III}&\multicolumn{1}{|c|}{Total}\\
\cline{2-13}1&2&3&4&5&6&7&8&9       &10&11&12&13\\
Case 1&0.5&1.5&2.1&4   &0(0)&1440(124)&1110(83)&2550(207)&0&2.6&3.6&6\\
Case 2&9.9&1.5&2.1&14   &3570(380)&1440(124)&1110(82)&6120(596)&10.3&2.6&3.6&17\\
Case 3&0.2&3.2&16.3&20 &0(0)&530(80)&2340(540)&2870(620)&0&2.9&6.1&8\\
Case 4&0.4&0.7&1.0&2    &0(0)&1340(28)&1050(18)&2390(46)&0&0.7&1.0&2\\
\cline{1-13}

\end{tabular*}
\end{center}
\end{table*}
 According to \cite{yun95} and \cite{lu06}, all progenitors of
SySs
pass through one of the three routes:\\
(i) Channel I: unstable Roche lobe overflow (RLOF) with formation of
a common
envelope;\\
(ii)Channel II: stable RLOF;\\
(iii)Channel III: formation of a white dwarf+red giant pair without RLOF. \\
Table \ref{tab:result} gives the results of SSs' and SSSs'
populations in SySs which have undergone channels I, II or III,
respectively.

In our model, SSs in SySs result from strong symbiotic novae. As
Table 2 shows, the Galactic occurrence rate of SSs in SySs is from
about 2 (case 4) to 20 (case 3) $\rm yr^{-1}$. Channels I, II and
III can produce SSs. According to \cite{yar05}, the typical duration
of mass loss ($t_{\rm ml}$) for a strong outburst is about dozens of
days. Therefore, SSs are transitory X-ray sources. SSSs in SySs
result from weak symbiotic novae or steady hydrogen burning
`ordinary' SySs. For SSSs from the former, their number is between
46 (case 4) and 620 (case 3), and their occurrence rate is from 2
(case 4) to 17 $\rm yr^{-1}$ (case 2). For SSSs from the latter,
their formation rate [between 0.01 (cases 1, 3 and 4) and 0.03 (case
2) $\rm yr^{-1}$] is very low and is not shown in Table
\ref{tab:result}. However, the lifetime of SSSs from steady hydrogen
burning is very long, which results in large numbers [ between 2244
(case 4) and 5537 (case 2)]. Due to $\gamma$-algorithm which gives
wide binary separation after common envelope evolution, SSSs in case
2 can be produced by channel I. While, SSSs in other cases do not
undergo channel I.

As the above descriptions, the input physical parameters (the
algorithm of the common envelope, terminal velocity of the stellar
wind $v(\infty)$ and the critical ignition mass $\Delta M_{\rm
crit}^{\rm WD}$) have great effects on the population of SSs and
SSSs in SySs. For SSs, $v(\infty)$ is a key factor, and the Galactic
occurrence rate of SSs in case 3 are about 5 times of those in case
1. For SSSs from weak symbiotic novae, $\Delta M_{\rm crit}^{\rm
WD}$ affects their Galactic number and occurrence rate within a
factor of 4; for SSSs from `ordinary' SySs, the algorithm of the
common envelope affects their Galactic number and formation rate
within a factor of 2.4. A detailed analysis for the effects of these
parameters can be seen in \cite{lu06}.


According to \cite{lu06}, the Galactic numbers of SySs in the cases
with the same input physical parameters are 5400 in case 1 and 4300
in case 4. The number ratios of the X-ray sources including SSs and
SSSs to SySs are $\sim$ 50\% in case 1 and $\sim$ 60\% in case 4.
$v(\infty)$ in case 3 is unfavorable for the formation of stable
hydrogen burning `ordinary' SySs \citep{lu06}, the number ratio is
about 20\%. \cite{mur97} found that 60\% of the known Galactic SySs
are sufficiently X-ray bright to be detected. This result is
agreement with our result. On comparing our results with
observations, one should note that the number of X-ray sources
depends greatly on the lifetime of the thermonuclear runaways as
X-ray sources.

In addition, it is very difficult to detect SSs and SSSs. The main
reason is that dominant interaction with interstellar matter for
X-rays between 0.1 keV and 10 keV is photoelectric absorption, and
the cross section increases rapidly at lower energies, scaling
approximately as $E^{-3}$ \citep{dm90}. According to \cite{m94},
SSSs have to be closer than $\sim$ 2 kpc to the Sun in order to be
detected. This means that only about $\frac{4}{225}$ of SSSs can be
observed. There are 4 SSSs whose distances from the Sun are shorter
than 2.6 kpc in \cite{mur97}. Hence, the Galactic number of SSSs in
SySs is about 200 if 4 SSSs in SySs are a completed sample within
2.6 kpc from the Sun. However, the orbital motion, the magnetic
field and the gravitational influence of the companions  in SySs can
complicate the structure of circumstellar material. The nova ejecta
and winds from the cool components in SySs act as shielding agents.
Therefore, 4 SSSs are a uncompleted sample.  A detailed work on
X-ray spectrum absorptions by circumstellar material and
interstellar medium is very difficult and beyond the scope of this
work.


\subsection{Properties of X-ray sources}
In this subsection, we describe potentially observable physical
quantities of SSSs and SSs in SySs.
\begin{figure*}
\begin{tabular}{c@{\hspace{3pc}}c}
\includegraphics[totalheight=4.5in,width=3.2in,angle=-90]{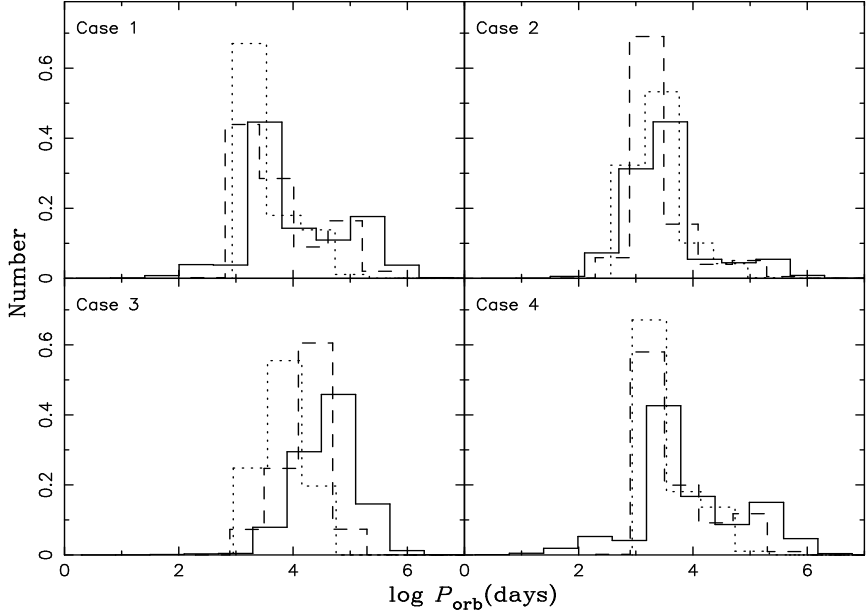}\\
\end{tabular}
\caption{Number distributions of X-ray sources in the Galactic SySs
as a function of orbital periods (all numbers are normalized to 1).
The solid, dashed and dotted lines represent SSs, SSSs from weak
symbiotic novae and from steady hydrogen burning `ordinary' SySs,
respectively.
            }
\label{fig:numporb}
\end{figure*}

Figure \ref{fig:numporb} gives the distributions of X-ray sources in
the Galactic SySs over orbital periods. As we can see from case 1,
SSSs have periods between 100 and 100000 days. SSSs from steady
hydrogen burning `ordinary' SySs and from weak symbiotic novae have
strong peaks at 7000 days and 11000 days, respectively. While SSs
have orbital periods from 10 to 1000000 days, with a typical peak at
36000 days. According to \cite{yun95} and \cite{lu06}, SySs via
channel I have short orbital periods, while they via channel III
have long orbital periods. As Table \ref{tab:result} shows, SSs can
be produced via channels I, II and III, but SSSs can be done via
channels II and III (except case 2). Therefore, SSs span wider range
of orbital periods than SSSs.

\begin{figure*}
\begin{tabular}{c@{\hspace{3pc}}c}
\includegraphics[totalheight=4.5in,width=3.2in,angle=-90]{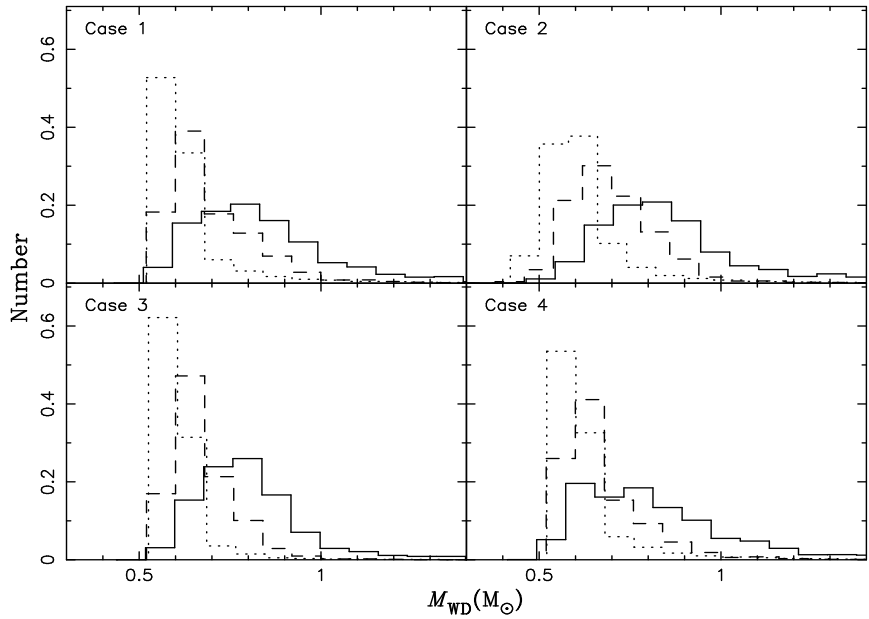}\\
\end{tabular}
\caption{Number distributions of X-ray sources in the Galactic SySs
as a function of WD mass (all numbers are normalized to 1). The
solid, dashed and dotted lines represent SSs, SSSs from weak
symbiotic novae and steady hydrogen burning `ordinary' SySs,
respectively.
            }
\label{fig:nummass}
\end{figure*}

Figure \ref{fig:nummass} shows the distributions of numbers  of the
modeled X-ray sources in SySs each as a function of WD mass. The
distribution for case 2 is wider than that in other cases. The main
reason is that $\gamma$ algorithm for common envelope can form SySs
with low-mass WD accretors \citep{lu06}. For all models, the
distributions have typical masses in the 0.5-0.8$M_\odot$ range, but
the peaks of the distributions are different from each type of X-ray
sources. For example, in case 1, SSs have a peak at about
0.7$M_\odot$. While, SSSs have a peak at about 0.6$M_\odot$, which
is in agreement with the results in \citet{yun96}. According to
\cite{lu06} and \cite{yar05}, massive WD favors to produce strong
thermonuclear outburst. \cite{m03} gave the distribution of the
measured masses of the WD in SySs. The peak is at about 0.5
$M_\odot$ regardless of \cite{m03} estimating low limits
\citep{lu06}. \citet{i03} showed average mass of WD in symbiotic
novae systems which are SSs is 0.73$M_\odot$.  It looks likely that
WD accretors in SSs have masses slightly larger than those in SSSs.
Our results are compatible with the observations.

\begin{figure*}
\begin{tabular}{c@{\hspace{3pc}}c}
\includegraphics[totalheight=4.5in,width=4.0in,angle=-90]{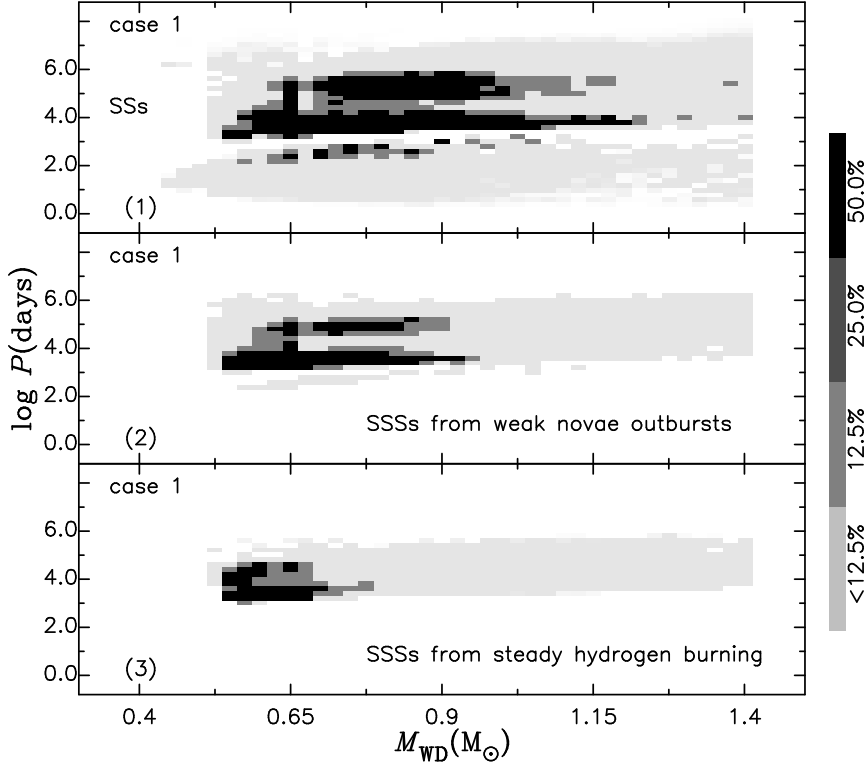}\\
\end{tabular}
\caption{Gray-scale maps of WD's mass vs. orbital
          period distributions
            for SSs and SSSs in SySs for case 1. The gradations of gray-scale
            correspond to the regions where the number density of systems is,
            respectively,  within 1 -- 1/2,
            1/2 -- 1/4, 1/4 -- 1/8, 1/8 -- 0 of the maximum of
             ${{{\partial^2{N}}\over{\partial {\log P}}{\partial {M_{\rm WD}}}}}$,
             and blank regions do not contain any stars.
            }
\label{fig:mp}
\end{figure*}

Figure \ref{fig:mp} shows the distributions of SSs and SSSs in case
1, in the `` WD's mass -- orbital period'' plane. The ranges of
orbital periods and WD's masses in SSs are wider than those in SSSs.
The main reason is that producing SSSs needs higher mass accretion
rates and the critical ignition mass $\Delta M_{\rm crit}^{\rm WD}$
than producing SSs. As the top panel in Figure \ref{fig:mp} shows,
the distribution in SSs is cut into two regions. The low region
represents these SSs which have undergone channel I, while the top
region is for SSs which have undergone channels II and III.

\begin{figure*}
\begin{tabular}{c@{\hspace{3pc}}c}
\includegraphics[totalheight=4.5in,width=3.2in,angle=-90]{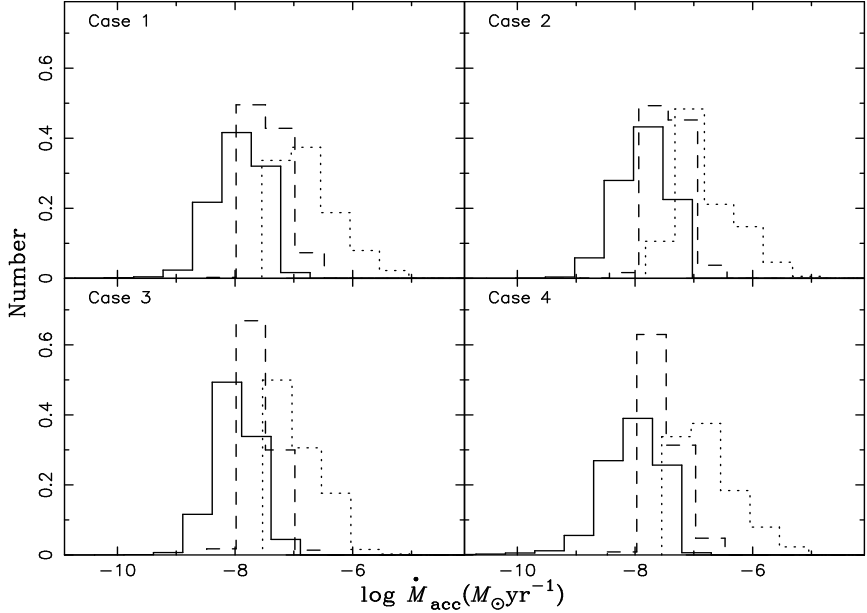}\\
\end{tabular}
\caption{Number distributions of X-ray sources in the Galactic SySs
as a function of the mass accretion rate of WD (all numbers are
normalized to 1). The solid, dashed and dotted lines represent SSs,
SSSs from weak symbiotic novae and from steady hydrogen burning
`ordinary' SySs, respectively.
            }
\label{fig:numdmdt}
\end{figure*}

The distributions of numbers of X-ray sources in the Galactic SySs
over the mass accretion rate of WD, $\dot{M}_{\rm acc}$, is given in
Figure \ref{fig:numdmdt}. According to our model, $\dot{M}_{\rm
acc}$ determines the strength of the hydrogen burning. For low
$\dot{M}_{\rm acc}$, the accreting WD experiences the strong
thermonuclear outburst which results in SSs. For high $\dot{M}_{\rm
acc}$, the accreting WD can undergo stable hydrogen burning.
Therefore, the peaks of magnitude of $\dot{M}_{\rm acc}$ in Figure
\ref{fig:numdmdt} are $10^{-9}$, $10^{-8}$ and $10^{-7}$$M_\odot \rm
yr^{-1}$ for SSs, SSSs from weak symbiotic novae and from `ordinary'
SySs, respectively. In \cite{i03}, theoretically estimated
$\dot{M}_{\rm acc}$ of five SSs is between $10^{-9}$ and $10^{-8}
M_\odot\rm yr^{-1}$, which is in good agreement with our results.

\section{Conclusions}
\label{conc}Assuming that SSs in SySs result from the violent
collisions of the stellar winds during strong symbiotic novae, and
SSSs from weak symbiotic novae or steady hydrogen burning SySs, we
investigate the Galactic population of SSs and SSSs in SySs. The
Galactic occurrence rates of SSs and SSSs are from $\sim$ 2 to 20
$\rm yr^{-1}$, and $\sim$ 2 to 17 $\rm yr^{-1}$, respectively. The
numbers of X-ray sources in SySs range from 2390 to 6120. The
percentage of the X-ray sources in all SySs is between $\sim$ 20\%
and 60\%, which is in agreement with the observations.

In the present paper we do not model the X-ray spectra emissions of
SSs and SSSs, and do not consider the interstellar medium's
absorptions for X-ray spectra. Our work only shows some primary
results. It is necessary for a detailed model of SSs and SSSs to
simulate X-ray spectra emissions and interstellar medium's
absorptions. We will consider them in the next paper.

\end{document}